\documentclass[twocolumn,showpacs,prl,superscriptaddress]{revtex4}

\usepackage{amsmath}
\usepackage{amssymb}
\usepackage{epsfig}
\usepackage{graphicx} 

\begin{document}

\title{A phonon laser in ultra-cold matter}

\author{J.T. Mendon\c{c}a}
\email{titomend@ist.utl.pt}
\affiliation{IPFN and CFIF, Instituto Superior T\'ecnico, 1049-001 Lisboa, Portugal}

\author{G. Brodin}
\affiliation{Department of Physics, Umea University, SE-901 87 Umea, Sweden}

\author{M. Marklund}
\affiliation{Department of Physics, Umea University, SE-901 87 Umea, Sweden}

\author{H. Ter\c{c}as}
\affiliation{IPFN and CFIF, Instituto Superior T\'ecnico, 1049-001 Lisboa, Portugal}


\begin{abstract}
We show the possible excitation of a phonon laser instability in an ultra-cold atomic gas confined in a magneto-optical trap. Such an effect results from a negative Landau damping of the collective density perturbations in the gas, leading to the coherent emission of phonons. This laser instability can be driven by a blue-detuned laser superimposed to the usual red-detuning laser beams which usually provide the cooling mechanism. Threshold conditions, instability growth rates and saturation levels are derived. This work generalizes, on theoretical grounds, the recent results obtained with single ion phonon laser, to an ultra-cold atomic gas, where real phonons can be excited. Future phonon lasers could thus adequately be called phasers.  
\end{abstract}

\maketitle

The last decade the concept of  phonon laser has been  studied in several different systems,  e.g.  quantum wells \cite{lozovik},  ions \cite{wallen}, nanomechanics \cite{barga}, and nanomagnets \cite{chud}. Furthermore,  very recently the possibility to create a Doppler instability and create a phonon laser with a single trapped atom or ion has been considered \cite{kaplan,vahala}.
Here we show that it is possible to generalized the single trapped atom configuration to the case of an large ensemble of atoms, thus enabling collective phonon excitations. In particular, the excitation of a phonon laser instability in an ultra-cold atomic gas confined in a magneto-optical trap (MOT) is described. Such an effect results from a negative Landau damping of the collective density perturbations in the gas, leading to the coherent emission of phonons. In contrast with the single ion case, here the phonon frequency is not determined by the single atom oscillation frequency due to the parabolic confining potential, but by the boundary conditions of the internal oscillations of the gas. It therefore corresponds to a collective oscillation, instead of a single particle effect. In that respect, it is closer to the paradigm of an optical laser, where the photon modes correspond to internal vibrations of the optical cavity, thus making the name \textit{phaser} a natural choice for such collective modes. The acoustic oscillations can also, in principle, be coupled to the outside world, by mechanical or electromagnetic means, thus providing a source of coherent acoustic radiation.

Collective processes in ultra-cold gas clouds have been considered by several authors in recent years. These processes result from the existence of an effective atomic charge in the gas cloud \cite{walker}, which create collective forces of the Coulomb type \cite{pruvost}. It was also shown that plasma-acoustic waves can be excited in the gas, which are similar to the usual sound waves but with a lower frequency cut-off \cite{mendkaiser}. These waves can provide the physical support of our laser instability.

Like in the case of a single trapped ion, the collective laser instability can be driven by a blue-detuned laser, superimposed to the usual red-detuned laser beams which usually provide the cooling mechanism. But in contrast with the single ion phonon laser, real phonons are excited in the medium and can eventually be coupled to an exterior system, just like photons produced in current optical laser cavities.

The dynamical behavior of the ultra-cold atomic gas will be described by a wave kinetic equation, where the atom recoil effects due to the emission and absorption of both photons and plasmons are retained \cite{mendnovo}. This equation corresponds to a straightforward generalization of well known quantum kinetic equations \cite{stenholm,cohen}, which include a collective (or mean-field) potential. The instability growth rates are derived, in the linear regime. Saturation of the phonon laser instability is also described with the help of a quasi-linear version of the wave kinetic equation.  Threshold conditions for the occurrence of the phonon laser instability are also established.
This work therefore extends a recently published concept of single ion phonon laser \cite{vahala} into the collective regime, by considering an ensemble of ultra-cold atoms confined in a MOT. In contrast to previous results, real phonons can be excited and eventually coupled to external environments. 

The starting point for our analysis is provided by the quantum wave kinetic equation for the ultra-cold gas \cite{mendnovo,cohen,stenholm}. This equation describes the space and time evolution of the Wigner function $W ({\mathbf r}, {\mathbf v}, t)$ associated with the kinetic state of the atoms in the cloud, where ${\mathbf r}$ and ${\mathbf v}$ are the position and the velocity of the atomic centre-of-mass. It is well known that this quantity provides the quantum analogue to the classical representation of the atom centre-of-mass in phase space. It is normally referred  to as a quasi-probability distribution since it can take negative values. Here we retain the exact quantum description of the centre-of-mass state of motion, which means that all the recoil effects associated with the atom interaction with electromagnetic and acoustic fields are taken into account. We will see that recoil is an essential ingredient of our phonon laser effect, but the classical limit will also be discussed. The evolution equation for the quantum quasi-probability distribution is 
\begin{eqnarray}
\frac{d W}{d t} = - i f (W) + \tilde{g}  \left[ W ( {\mathbf k}_b )^{(-)} - W ( {\mathbf k}_b )^{(+)} \right] 
\nonumber \\ 
+ \sum_n g_n \left[ W ( {\mathbf k}_n )^{(-)} - W ( {\mathbf k}_n )^{(+)} \right]  ,
\label{eq:1} 
\end{eqnarray}
where the total time derivative is $d/dt \equiv ( \partial / \partial t+  {\mathbf v} \cdot \nabla)$. This equation states that the changes in the  distribution $W ({\mathbf r}, {\mathbf v}, t)$, are due to three different factors. The first is the external force term, associated with the collective mean-field potential as defined by
\begin{equation}
f (W) = \frac{1}{\hbar}  \int V ({\mathbf k}) \left[ W ( {\mathbf k} )^{(-)} - W ( {\mathbf k} )^{(+)} \right] e^{ i {\mathbf k} \cdot {\mathbf r}} \frac{d {\mathbf k}}{(2 \pi)^3}  ,
\label{eq:2} 
\end{equation}
where
$W ( {\mathbf k} )^{(\pm)} \equiv W ({\mathbf v} \pm \hbar {\mathbf k} / 2 M )$, with $M$ the mass of the atom, and we have used a spatial Fourier decomposition in the integral in (\ref{eq:2}). This collective force describes the atom-atom interactions inside the gas and is determined by the potential $V$, formally analogous to an electrostatic potential, governed by the Poisson equation
\begin{equation}
\nabla^2 V = - Q n \equiv - Q \int W ({\mathbf r}, {\mathbf v}, t) d {\mathbf v} .
\label{eq:3} \end{equation}
Here $Q$ determines the effective charge of the atom \cite{walker,pruvost,mendkaiser}, as will be discussed later, and depends on the intensity of the incident cooling and pumping laser beams. The second source in Eq.\ (\ref{eq:1}) is due to the red-detuned cooling laser beams (normally six in number) with electric fields of the form ${\mathbf E} ({\mathbf r}, t) = {\mathbf E}_n \exp (i {\mathbf k}_n \cdot {\mathbf r} - i \omega_a t)$, corresponding to the same frequency $\omega_a$, which is nearly equal but slightly lower than the frequency of the atomic transition used for laser cooling, and wavenumbers $| {\mathbf k}_n | = k_a = \omega_a / c$. In equation (\ref{eq:1}), the coupling coefficients $g_n$ are associated with the red-detuned laser beams. Finally the third source term in Eq.\ (\ref{eq:1}), is due to the blue-detuned laser pump field, ${\mathbf E} ({\mathbf r}, t) = {\mathbf E}_b \exp (i {\mathbf k}_b \cdot {\mathbf r} - i \omega_b t)$,  with frequency $\omega_b > \omega_a$. This pump field will produce a population inversion in the center of mass velocity state, leading to an instability of the acoustic waves in the ultra-cold gas. We already now stress that these acoustic waves have some special features which make them clearly distinct from common acoustic oscillations. This will become apparent later. 

It is obvious from equation (\ref{eq:1}) that the first term couples the atomic states with the spectrum of possible low frequency oscillations of the collective potential $V ({\mathbf k})$, with $|{\mathbf k}| \ll {\mathbf k}_b| \simeq | {\mathbf k}_a|$, and the second term couples the atoms with the blue-detuned pumping field. These two terms can describe the atomic recoil associated with the exchange of momentum with both photons (second term) and phonons (first term).
The radiation coupling coefficient, which is the mediator of photon recoil, is given by the usual expression $\tilde{g} = - ( E_0 / \hbar) \Im ( d_{21} \tilde{\rho}_{21})$, where $d_{21}$ is the dipole matrix element of the radiative transition between the internal atomic states $1$ and $2$, and $\tilde{\rho}_{21}$ is the density matrix element, assumed nearly at equilibrium with the radiation field. In the above kinetic equation we have neglected the confining potential, which is assumed to be parabolic. This is justified by the fact that the wavelength of the collective oscillations are assumed much shorter than the size of the confining potential well. Therefore, as a first approximation, the medium can be assumed as infinite. The role of the boundary conditions will be discussed at the end.

In order to simplify the above description we consider an one-dimensional model, by assuming ${\mathbf v} = u ( {\mathbf k} / k) + {\mathbf v}_\perp$, and integrate over the perpendicular velocities with respect to the pump laser beam. In the low frequency spectrum of the collective potential $V ({\mathbf k})$, we retain spectral components  propagating in the same direction, ${\mathbf k} \parallel {\mathbf k}_b$. We define the  quasi-distribution for the parallel velocities as $G (u) = \int W (u, {\mathbf v}_\perp) d {\mathbf v}_\perp$. The resulting evolution equation is
\begin{equation}
\frac{d G}{d t} = - i f (G) + \tilde{g}  \left[ G ( k_b )^{(-)} - G ( k_b )^{(+)} \right] + \nu \left[ G - G_0 \right]  
\label{eq:4} \end{equation}
where $\nu$ is a phenomenological viscosity coefficient. In reality, this quantity depends on the angle between the pumping and the cooling beams. But, for order of magnitude estimates, we can use the  the viscosity coefficient, as defined in the absence of the pumping laser beam \cite{pruvost,mendkaiser}. The distribution $G_0 (u)$ is that of the laser cooled gas, as obtained in the absence of the pumping laser. This last term in (\ref{eq:4}) plays the role of spontaneous emission by depopulating the higher velocity states driven by $\tilde{g}$, as shown below.

Three physical processes are assumed to occur in parallel, as described in figure 1. We can see a very strong resemblance with the usual three level laser model. First, the red-detuned laser beams, will create a very low temperature quasi-distribution $G (u)$, which corresponds to our ground state. Second, the blue-detuned pump laser field will excite high velocity atomic states around the velocity $u + \hbar k_b / M$, due to the absorption of photon momentum, therefore populating the upper level for the phonon laser transition. This high velocity distribution corresponds to a kind of particle beam, thus creating a population inversion in the centre-of-mass states. It will then lead to a negative Landau damping of acoustic-like oscillations with frequency $\omega$ and wavenumber $k$, which will result in the coherent emission of phonons. Due to phonon emission, a third and intermediate velocity state will be populated. The temporal evolution of the phonon field will be dictated by a complex frequency $\omega = \omega_r + i (\gamma_k - \nu / 2)$. 
\begin{figure}
\begin{center}
\epsfig{file=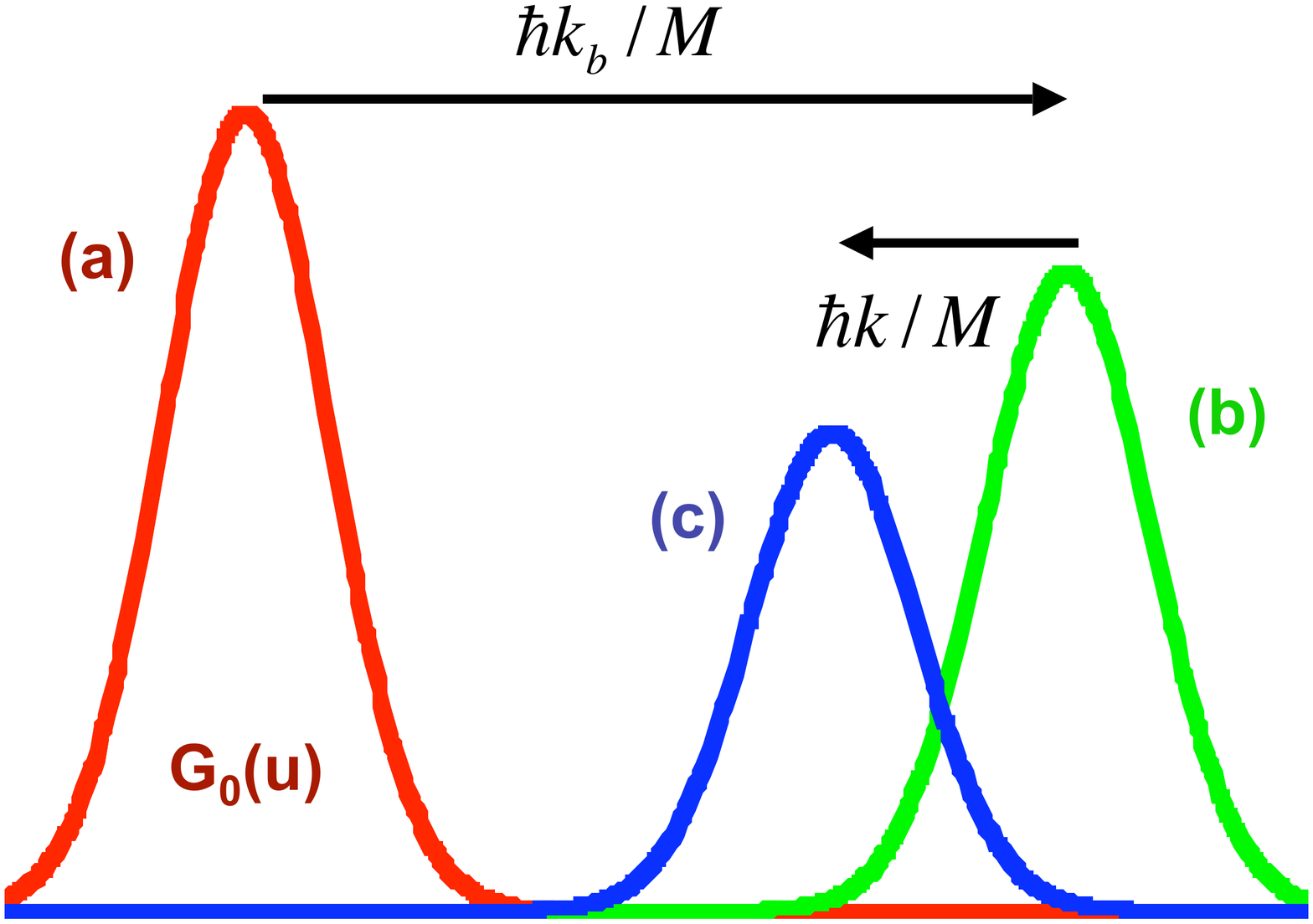,width=8.5cm,clip=}
\end{center}
\caption{{Phonon laser scheme: (a) The gas is cooled down by red-detuned laser beams with frequency $\omega_a$ and wavevector ${\mathbf k}_a$ into a velocity distribution $W_0 (v)$; (b) the ultra-cold gas is pumped by a blue-detuned laser with frequency $\omega_b$ and wavevector ${\mathbf k}_b$, therefore creating an inversion of population in velocity space; (c) Phonons with frequency $\omega$ and wavevector ${\mathbf k}$ are coherently emitted by the atoms, which decay into a lower kinetic energy state. A one-dimensional configuration is assumed for simplicity.}}
\label{fig:1}
\end{figure}
We assume that $G (u) = \bar{G} (u) + \tilde{G} (u)$, where $\bar{G} (u)$ is the equilibrium distribution, and $\tilde{G} (u)$ is the perturbation, which is assumed to evolve as $\exp (i{\mathbf k} \cdot {\mathbf k
r} - i \omega t)$. By linearizing equations (\ref{eq:1}) and (\ref{eq:3}), it is possible to show that the instability growth rates are determined by  \cite{mendnovo}
\begin{equation}
\gamma_k = -  \frac{\pi Q \omega_r}{2\hbar k^3} \left[ \bar{G}^{(-)} - \bar{G}^{(+)}  \right]_r
\label{eq:5} \end{equation}
where the population difference is calculated for the resonant parallel velocity $u_r = k / \omega_r$, and the acoustic oscillations obey the dispersion relation 
\begin{equation}
\omega_r^2 = \omega_p^2 + k^2 u_s^2 + \frac{\hbar^2 k^4}{4 M^2}
\label{eq:5b} \end{equation}
This dispersion relation describes a modified form of acoustic oscillations.
The first term on the right hand side of Eq.\ (\ref{eq:5b}) corresponds to a plasma frequency cut-off, the second term is the usual acoustic dispersion term, and the last term is a purely quantum correction due to the atom recoil. The plasma frequency $\omega_p$ and the ion sound speed $u_s$ are determined by
\begin{equation}
\omega_p = \sqrt{\frac{Q n_0}{M}} \; , \quad u_s^2 = \frac{1}{n_0} \int G (u) u^2 d u
\label{eq:5c} \end{equation}
where $n_0$ is the unperturbed gas density. By comparing this definition of the plasma frequency of the neutral gas with the usual definition valid for an electron-ion plasma, we conclude that the effective charge of the neutral atoms inside the ultra-cold gas is $\sqrt{\epsilon_0 Q}$, as first considered by \cite{walker}.
It is clear that we will have an acoustic wave growth if the inverse Landau damping is positive $\gamma_k > 0$, and if it is large enough to compensate for the wave losses, i.e.\ $\gamma_k > \nu$. We can then write the threshold condition as
\begin{equation}
\left[ \bar{G}^{(-)} - \bar{G}^{(+)}  \right]_r > \frac{2 \nu}{\pi} \frac{\hbar k^3}{Q \omega_r}
\label{eq:6} \end{equation}
It should be noticed here that such an instability remains in the  classical limit. In this limit, we can neglect the atom recoil due to emission or absorption of phonons, which allows us to develop $\bar{G}^{(\pm)}$, around $\bar{G} (u)$, as $ \bar{G}^{(\pm)} \simeq \bar{G} (u) \pm (\hbar k / M) (\partial \bar{G} / \partial u)$. The threshold condition now reads 
\begin{equation}
\left( \frac{\partial \bar{G}}{\partial u}  \right)_r > \frac{2 \nu}{\pi} \frac{M k^2}{Q \omega_r}
\label{eq:6b} \end{equation}
The population inversion is now represented by the derivative of the atom distribution at the resonant velocity $u_r = \omega_r / k$. The resemblance with a three level laser will be somewhat lost, but the physical principle stays the same. 
\begin{figure}
\begin{center}
\epsfig{file=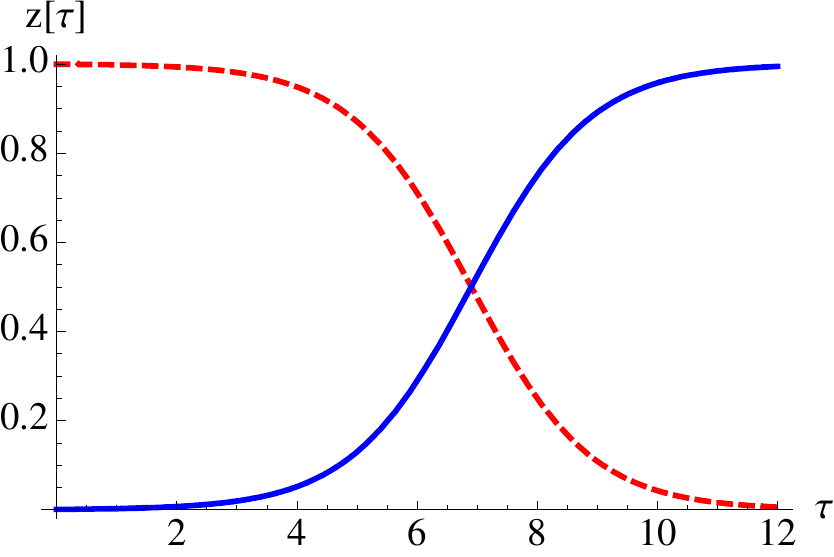,width=8.5cm,clip=}
\end{center}
\caption{{Saturation of the laser instability, as described by the quasi-linear equations. Temporal evolution of the normalized population inversion in velocity space $z (\tau) \equiv \Delta G (u, \tau) / \Delta G (u, 0)$, and the phonon amplitude square $y (\tau) \equiv | V_k (\tau )|^2$, as a function of the time variable $\tau$, for $a = 1$ and $\bar{\nu} = 1/2$. For illustration purposes, we take $y (0) = 0.001$}}
\label{fig:2}
\end{figure}
Next, we discuss the instability saturation. 
The acoustic wave growth will tend to decrease the population difference between the upper and lower kinetic energy levels. This process is approximately described by the quasi-linear equations \cite{mendnovo}
\begin{equation}
\frac{d \bar{G}}{d t}= \frac{| V ( {\mathbf k}, t)|^2}{\hbar^2} 
\left[ \bar{G} (u - \hbar k / M ) - \bar{G} ( u + \hbar k / M ) \right] 
\label{eq:7} \end{equation}
In order to be specific, we assume that the mean distribution $\bar{G} (u)$ is divided into three distinct regions of spectral interest, centered arround $u_0 = 0$, $u_1 = \hbar (k_b - k) / M$ and $u_2 = \hbar k_b / M$. We have, $\bar{G} ( u) = G_0 (u) + G_1 (u) + G_2 (u)$. For the extreme case of $G_j (u) = G_{j0} \delta (u - u_j)$, with $j = 0, 1, 2$, the above quasi-linear equation would be reduced to simple balance equations between the three velocity level populations. The resonant atom velocity defined in equation (\ref{eq:5}) for the inverse Landau damping will then be exactly equidistant from $u_1$ and $u_2$. Integrating over the parallel velocity, and defining the population difference $\delta G = \int [G_2 (u) - G_1 (u) ] d u$, for the atoms centered around the upper and intermediate parallel velocities $u_1$ and $u_2$, we can easily establish the evolution equation
\begin{equation}
\frac{d }{d t} \Delta G = -  \frac{| V ( {\mathbf k}, t)|^2}{\hbar^2} \Delta G
\label{eq:8} \end{equation}
On the other hand, the inverse Landau damping of the phonon mode will also change and, as a consequence, the phonon square amplitude (energy) will evolve according to 
\begin{equation}
\frac{d}{d t} | V_k ( t )|^2 = \left [ \frac{\pi Q \omega_r}{\hbar k^3} \delta G -  \nu \right] |V_k ( t ) |^2
\label{eq:9} \end{equation}
where $\gamma_k$ is the growth rate defined above. 
Introducing a new time variable, $\tau$, and a new viscosity coefficient $\bar{\nu}$, such that
\begin{equation}
\tau = ( \hbar k^3 / \pi Q \omega_r) \Delta G (0) t \; , \quad \bar{\nu} = (\tau / t) \nu
\label{eq:9b} \end{equation}
 we can rewrite these two equations as
\begin{equation}
\frac{d z}{d \tau} = - a z y \; , \quad \frac{d y}{d \tau} = (z -  \bar{\nu} ) y
\label{eq:10} \end{equation}
with $z \equiv \Delta G / \Delta G (0)$, $y \equiv | V_k |^2$ and $a = (\pi Q \omega_r / \hbar^3 k^3)$.These coupled evolution equations clearly exhibit the threshold condition $z (0) > 2  \bar{\nu}$, already stated in equation (\ref{eq:6}), and saturation for increasing values of time.This is illustrated in figure 2. We can see that the population difference decreases along time, due to the increase of the phonon coupling between the two velocity states around $u_1$ and $u_2$. As a consequence, the growth rate slows down, leading asymptotically to saturation. 

In our discussion we have only considered a single phonon mode, corresponding to a given wavenumber $k$. It is important to discuss the selection mechanisms identifying such a mode inside the phonon spectrum. Several different processes can lead to mode selection. First, we have spontaneous selection of the mode with the largest growth rate. Due to the exponential growth, such a mode can easily detach from the background noise. However, a more effective mechanism is due to the finite size of the gas cloud, which can be confined in a MOT. This leads to a discretization of the phonon mode spectrum, as recently discussed \cite{mendkaiser}, and the resulting discrete phonon modes with a well defined internal structure are called the Tonks-Dattner resonances of the nearly spherical gas. In a certain sense, this can be seen as the phonon equivalent of the laser modes in a typical laser cavity. The dominant phonon mode in an ultra-cold gas will be due to the Tonks-Dattner resonance mode with the largest inverse Landau damping. Recent experimental work suggests that such modes can be spontaneously excited in the absence of any pump laser beam \cite{hugo}. Typical experimental conditions are $10^{9} - 10^{10}$ confined atoms, and plasma frequency (the lower cut-off for Tonks-Dattner oscillations) of $100$ to $200 \,\mathrm{Hz}$. Finally, these modes can be excited by an external source, for example due to amplitude modulation of one of the incident laser beams, and the phonon laser system will operate as an amplifier. Such a phonon laser system, operating as an oscillator or as an amplifier, could adequately be called a phaser (instead of saser, as proposed by others).

In conclusion, we have shown that a phonon laser instability can be excited in the ultra-cold atomic cloud confined in a MOT, and pumped by a blue-detuned laser beam. Threshold conditions, linear growth rates and nonlinear saturation have been established. We have used a quantum kinetic description, which retains the atom recoil effects associated with the emission and absorption of both photons and phonons. This work generalizes recent proposals for a single atom phonon laser \cite{kaplan,vahala} to a large ensemble of atoms . The present phonon laser configuration is able to produce real phonons in the gas. Due to the acoustic cavity, the phonon modes are determined by the boundary conditions of the gas, in contrast to the single atom oscillation frequency studied in previous works.

\end{document}